\documentclass[]{spie}  

\usepackage[]{graphicx}
\def\titlefont{\LARGE\bf} 
\title{A quantum symmetric key cipher(Y-00) and key generation   (Quantum stream cipher-Part II)} 

\author{
Osamu Hirota\supit{a,b,c}, 
Kentaro Kato\supit{b,c}, 
Masaki Sohma\supit{a,b},
Masaru Fuse\supit{d}
\skiplinehalf
\supit{a}
Research Center for Quantum Information Science, Tamagawa University, 
 Tokyo, Japan\\
\supit{b}
Japan Science and Technology, CREST project\\
\supit{c}
21st century COE program, Chuo University, Tokyo, Japan\\
\supit{d}
Panasonic(Matsushita Electric Industrial CO., Ltd),
 Osaka, Japan
}

\authorinfo{
Further author information:\\
Osamu Hirota: address; 6-1-1, tamagawa-gakuen, machida-city, Tokyo, Japan. E-mail, hirota@lab.tamagawa.ac.jp, Telephone, +81-42-739-8674.\\
}

\begin{document} 
  \maketitle 

\begin{abstract}
What obstructs the realization of useful quantum cryptography
is single photon scheme, or entanglement which is not applicable to the current infrastructure of
 optical communication network.  
We are concerned with the following question:
Can we realize the information theoretically secure symmetric key cipher under "the finite secret key" based on quantum-optical communications?
A role of quantum information theory is to give an answer for such a question.
As an answer for the question,  a new quantum cryptography was proposed by H.P.Yuen, which  can realize a secure symmetric key cipher  with high speeds(Gbps) and for long distance(1000 Km).
Although some researchers claim that Yuen protocol(Y-00) is equivalent to the classical cryptography, they are all mistaken. Indeed it has no classical analogue, and also  provides a generalization even in the conventional cryptography. 

At present, it is proved that a basic model of Y-00 has at least the security such as $H(X|Y_E)=H(K|Y_E)=H(K)$, $H(K|Y_E,X)\sim 0$ under the average photon number per signal light pulse:$<n> \sim 10000$. Towards our final goal, in this paper, we clarify a role of classical randomness(secret key) and quantum randomness in Y-00, and give a rigorous quantum mechanical interpretation of the security, showing an analysis of quantum collective attack.
\end{abstract}


\section{INTRODUCTION}
In 1984, Bennett and Brassard[1] gave an impact to the field of cryptography by showing a new protocol based on quantum mechanics so called BB-84.
It is well known that they opened not only a new scientific subject, but also  called  attention of quantum information scientists and theorists of cryptography to information theoretic cipher again. We are concerned with further development of information theoretic cipher based on quantum cryptography initiated by BB-84. 

Let us describe, first, a story of unconditional security based on the standard text book.  
Many works on  protocol with unconditional security  have been already discussed in journals of information theory.
The definition of  perfect secrecy is $I(X;C)=0$. It  means that the plaintext $X$ and the cipher text $C$ as a function of $X$ and a secret key $K$ should be statistically independent. However, in this discussion  Eve has access to precisely the same information as the legitimate users. 
As a result,  the condition: $H(X)\le H(K)$ is required for the perfect secrecy. 
It means that perfect secrecy is achieved only when the secret key is at least as long as the plaintext message. However, such a pessimism may be solved by introducing the modified Shannon's model such that Eve cannot receive precisely the same information as Bob. In general, there are conditions to realize the perfect secrecy. If we have no condition for it even under any physical laws, then the scheme is called unconditionally secure.

First, in order to seek a perfect secrecy, an information theoretic cipher for noisy channel was discussed.
For general discussions, a channel model of secure communication  by  conventional information theory is defined. 
That is, we have two channels. One is 
a channel from Alice to Bob, and the other is  from Alice to Eve. 
Let $X,Y,Z$ be random variables of Alice, Bob, and Eve, respectively. Channels are completely specified by the conditional probability $P(Y, Z | X)$.
A confidential communication without key under such a concept was given by Wyner and et al, and it is called information theoretic cipher.
As an example, Wyner[2] clarified a fact that one can realize a scheme with unconditional security in which Eve is assumed 
to receive signals from Alice over a channel that is noisier than 
the legitimate users. Subsequently Csisz$\acute{\rm}$ar-K$\ddot{\rm o}$rner[3] generalized Wyner's result. These schemes allow us to realize one way communication with confidential 
message $X$ without initial key, and also distribution of key $K$. 
However, the assumption that Eve's channel is worse than the legitimate channel is also unrealistic. 
When the discussion is devoted only to  the problem of  key distribution, Maurer[4] pointed out that this assumption is not needed if the legitimate users can communicate over insecure but authenticated public channel like BB-84.
Unfortunately, the efficiency of communication is not so good. 
In addition, in such information 
theoretical results, the problem of finding actual encodable and 
decodable codes that perform in a particular situation was remained.
Thus the unconditionally secure key distribution  requires neither  single photon communication nor even quantum phenomena. These results verified that the key distribution with unconditional security is not proper issue of quantum theory. So one can realize unconditionally secure scheme by classical communication systems if there exist unavoidable 
noise such as "thermal noise in free space" and so on. In addition, $H(X)\le H(K)$ is not essential.  
However, in principle, if any noises in the system are removable, then one exactly needs quantum mechanical law to realize unconditional security.
 In any situations, the most important concept is of an advantage distillation or advantage creation.

On the other hand, there is cryptography based on another criterion, so called "computational complexity based security".
Symmetric key cipher and public key cipher belong to this category.
In the conventional cryptography, there is no encryption scheme with provable security in the sense of information theoretic security for ciphertext only attack and known plaintext attack on key, because the security comes from only key uncertainty.
One of the methods to provide "provable security" may be quantum cryptography.
There are many directions for further development, but it depends on personal fancy, for example, 
\begin{itemize}
\item[\rm(i)]  quantum key generation with unconditional security
\item[\rm(ii)] direct encryption by quantum key distribution and one time pad
\item[\rm(iii)] quantum symmetric key cipher with information theoretic security
\end{itemize}
Even in any purpose, they will be not accepted in the real world if they have no efficiency. So the ultimate new cryptography should satisfy both {\bf security and efficiency requirements}. 
We are concerned with third category under the following conditions:
\begin{itemize}
\item[\rm(i)]  Eve has the computer with unlimited computation power
\item[\rm(ii)] Eve has unlimited physical resource
\end{itemize}
One candidate for the realization is to use noisy channel governed by unavoidable quantum noise. Yuen raised a question:
"Is it possible to create a quantum system with current technology that could provide a communication in which always Bob's error probability is superior to that of Eve?", and 
in 2000, he gave a protocol(Y-00) to realize it as a positive answer[5,6], in which he proposed a new scheme  with  $M$-ary quantum state modulation.  In fact, 
this provides a new basis to communication with confidential message even in the conventional cryptography. 

In this paper, to reveal an excellent  potential of Y-00,  we shall give a rigorous interpretation of security principle of Y-00.

\section{Basis of quantum communication for quantum cryptography}
In any quantum cryptography such as BB-84, B-92, E-91, and Y-00,  the information is classical bit, not quantum information. That is, the information is true random number for key distribution, and plaintext for direct encryption.
The essential assumption in quantum communication for classical information is that quantum states are known for the legitimate users. So classical bits are mapped into a set of known quantum states. 
They are transmitted passing through a completely positive map(cp-map), and  discriminated by quantum measurement process described by positive operator valued measure(POVM). Then, a receiver gets classical bits as information by measurements. Such a model is called Helstrom/Holevo/Yuen formalism for quantum communication[7,8].
Let us give a brief introduction.
An information source and output in the quantum communication model are described by a density operator for ensemble of quantum states which conveys classical information as follows:
\begin{equation}
\rho_{Tin} =\sum p_i \rho_i,\quad \rho_{Tout}= \sum p_i \epsilon(\rho_i),
\end{equation}
where $i$ is an index corresponding to symbol as classical information, and $\epsilon$ is a cp-map.
The discrimination among quantum states at the output of the channel is described by POVM.
\begin{equation}
\Pi_j \ge 0, \quad \sum \Pi_j =I,
\end{equation}
where $I$ is an unit operator.
Then a conditional probability for each trial of the measurement is given by 
\begin{equation}
P(j|i)=Tr \epsilon(\rho_i) \Pi_j.
\end{equation}
The minimization problem of the average error probability based on the above equation is called quantum detection theory, which is a fundamental formalism in quantum information science.
\begin{equation}
P_e =\min_{\Pi} \{1-\sum p_i Tr\epsilon(\rho_i )\Pi_i\}
\end{equation}
The complete theory has been given by Helstrom/Holevo/Yuen. 
As a result, we have [7,8]
\\
{\bf Theorem 1}: 
{\it Signals with non-orthogonal states cannot be 
distinguished without error and  optimum lower bounds for error rate exist.}\\
\\
This means that if we assign non-orthogonal states for bit values 1 and 0,
then one cannot distinguish 1 and 0 without error.
When the error  probability is 1/2, there is no way to distinguish them. The most important cp-map(communication channel) in the real world is energy loss channel with $20 \sim 100 dB$ loss. A selection of  input quantum states for the channel is one of interesting problems in quantum communications, but we have the following result[9].
\\
{\bf Theorem 2}: 
{\it The input state which keeps pure state passing through energy loss channel is only coherent state.}\\
\\
So we can understand that a desirable state is coherent state.
In fact, the ultimate information transmission for such a model is discussed by Holevo capacity theory, and the ultimate capacity formula is well known Holevo/Schumacher/Westmorland theorem[10] as channel coding theorem. The Holevo capacity for energy loss channel, which gives the ultimate efficiency for quantum communication by quantum states, has the following quantitative property[11,12].
\begin{equation}
C_H=\chi_H(coherent)\ge \chi_H(squeezed:1mode)\gg \chi_H(squeezed:2modes)  \gg \chi_H(1photon)
\end{equation}
where $\chi_H$ is Holevo function.
So  any states except for coherent state are out of the scope for communications at present. Thus single photon state, squeezed state, entangled state, and so on are not useful in the real infrastructure of communication.

On the other hand, in the quantum communication model for quantum cryptography, we have to consider two channels of Alice to Bob, and Alice to Eve.
Let us describe them by $\epsilon_{AB}$, $\epsilon_{AE}$.
In general, $\epsilon_{AE}$ is ideal channel while $\epsilon_{AB}$ is noisy channel. The basic performance of cryptography is to prevent a leak of secret information from channel of legitimate users.
In a physical cryptography like quantum cryptography, one may take a method to eliminate Eve's information obtained by her measurement from $\epsilon_{AE}$. In order to realize such a situation, one needs "advantage distillation" under the ultimate physical law.
 It means that the defects of Bob can be got rid of by some processing, while the performance of Eve, who has the unlimited power of computer and physical resources, is superior than that of Bob in the original situation.

Thus, when we  take into account two criteria: efficiency and security  as requirement to quantum communication for quantum cryptography, the most preferable state is mesoscopic coherent state.
Even in BB-84, there are many proposals for realization based on mesoscopic coherent state[13] which are very welcome.

\section{Yuen protocol(Y-00) by coherent state}
\subsection{Unification of symmetric key cipher and information theoretic cipher}
Our purpose is to devise a cipher with an information theoretic security, which prevents Eve from finding the unique data or key even with unlimited computation power.

A symmetric key cipher is a scheme that Alice and Bob share a secret key. A block cipher and a part of stream ciphers belong to this category. However, they are in principle insecure, because the security is given by only key uncertainty. In addition, 
a  secure communication by "one way scheme" in information theoretic cipher requires a situation  that the channel between Alice and Eve is very noisy, but that of Alice and Bob is  a normal communication. It is unrealistic.
On the other hand, there are generalized stream ciphers which are designed by information theoretic approach and randomized approach proposed by C.Schnorr, Cachin, Maurer, and Diffie[14]. 
However, they have some conditions. For examples, C.Schnorr's stream cipher is that Eve can access only limited number of ciphertexts, Cachin-Maurer's cipher works under the assumption of limited memory capacity[14]. 
These generalized stream ciphers give a hint to devise a symmetric key cipher with information theoretic security, because 
our problem is to realize secure communication by a symmetric key cipher. So our purpose may be achieved if and only if one can unify  symmetric key cipher(especially generalized stream cipher) and information theoretic cipher.

We shall show how this unification is done.
According to a quantum detection theory we have the following properties for average error probability:
\begin{equation}
P_e(BP) < P_e(BM), \quad P_e(BP) < P_e(MP)
\end{equation}
where $BP$, $BM$, $MP$ mean binary pure state, binary mixed state, and $M$-ary pure state, respectively. The problem is  how to apply the above principle of quantum detection theory to cryptography.
Yuen proposed a protocol which combines  a  shared secret  key  for the legitimate users and  specific quantum state modulation scheme. A main idea of this protocol is the explicit use of a shared secret key and physical nature of noise for cryptographic objective of secure communication and key generation. {\it This is called initial shared key advantage} in noisy channel. 
By this advantage, the legitimate users can establish "advantage distillation" or "advantage creation" under the finite size for any parameters of the protocol in noisy channel. 
As a result, one can see "a basic principle to guarantee the security " as follows[6]:\\

{\bf Principle of security }: 
{\it The optimum quantum measurements with key and without key have different performance.}\\

Unknown key corresponds to classical randomness. The security of the conventional symmetric key cipher comes from this classical randomness. However, in Yuen protocol, a classical randomness is used to make a difference of the performance of quantum measurements. It means that if Eve does not know the key, then the quantum limitation of her measurement is enhanced by classical randomness. As a result, Eve has to search the data or key based on her measurement results with unavoidable error.
Thus, in general, although a symmetric key cipher belongs to the class of ciphers of computational complexity based security, by this principle one can realize a symmetric key cipher with information theoretic security. It will be instructive to compare the concept of Yuen, Schnorr, and Cachin.
\begin{itemize}
\item[\rm(i)]  Yuen: Eve's data on plaintext or key has unavoidable error without any conditions for Eve.
\item[\rm(ii)] Schnorr: It works under the condition that Eve can get only limited number of exact cihertext. 
\item[\rm(iii)] Cachin-Maurer: It works under the condition that Eve can use only limited memory capacity. 
\end{itemize}

For explanation of this principle, Yuen gave a simple example without any design as follows:
If Eve wants to know some information on the data bits, she has to measure the signals by any instrument. In  Y-00 of the original model, the problem of the measurement ability reduces to the comparison with optimum binary quantum measurement and optimum phase measurement. Since Eve does not know $K$, she needs to make the phase estimation in order to identify data $X$ for all possible basis selection from the running key.
According to the quantum detection theory, when Eve and Bob have the ultimate ability(ultimate receiver devices, and so on), their error probabilities are shown for binary signals with key and without key as follows[6]: 
\begin{equation}
{P_e}^B \sim \exp(-4S) \quad {\rm vs} \quad {P_e}^E \sim \exp(-2S)
\end{equation}
where $S=<n>$ is signal energy. Thus the error probability of Bob is smaller than that of Eve. This fact gives an advantage distillation under the ultimate physical law, so it leads to unconditionally secure key generation for the any key length of the initial key, and also it gives a basis for  information theoretically secure direct encryption.
The above example is not a scheme what we use as a practical quantum cryptography. Only it shows a principle. For practical use, we need several additional contrivances. 
The essential problem is that how to extend the above principle towards practical quantum cryptography.
Yuen has suggested the following directions, showing the more general unification theory as KCQ[6].
\begin{itemize}
\item[\rm(i)]  Direct encryption: quantum stream cipher( or $\alpha\eta$ scheme) as a randomized cipher by quantum noise. 
\item[\rm(ii)] Key generation: generalized Y-00 based on coherent pulse position modulation and so on.
\end{itemize}

The key point is a role of the classical randomness for quantum cryptography. A first idea for use of classical randomness was proposed as follows. We assume that Alice and Bob share a secret  key $K$. The key is stretched by a pseudo random number generator to $K'$.
The data bit is modulated by $M$-ary keying driven by random decimal number generated from the block :$K'/log M=\bar{K}'=(k_1,k_2,\dots)$ of pseudo random number with the seed key $K$. 
The $M$-ary keying has $M$ different basis based on 2$M$ coherent states. So the data bit is mapped into one of 2$M$ coherent states randomly, but of course its modulation map has a definite relationship based on key, which is opened. This is a fundamental structure of Y-00[15]. We shall describe a feature of Y-00 by the most simple way in the following sections.

\subsection{Quantum stream cipher and the security}
An application of Y-00 is, first,  direct data encryption like a stream cipher in the conventional cryptography. 
We call the symmetric key cipher based on Y-00 protocol "quantum stream cipher" or $\alpha\eta$ scheme[15,16,17].
Here it is reasonable that we  employ different security criteria for  direct encryption and  key generation.
For  direct encryption, the criteria are given as follows.
\begin{itemize}
\item[\rm(i)]  Ciphertext-only attack(CTOA) on data and on key: To get plaintext or key, Eve knows only the ciphertext from her measurement.
\item[\rm(ii)] Known/chosen plaintext attack(KTA): To get key, Eve  inserts  her known or chosen plaintext data into modulation system( for example, inserts all 0 sequence as plaintext in a period). 
Then Eve tries to determine key from input-output. Using the key, Eve can determine the data from the ciphertext.
\item[\rm(iii)] Repetition attack: Since the secret key is fixed, it has a period. Eve can apply CTOA and KTA over many periods when the key is reuse.
\end{itemize}

In order to use effectively the principle of security, quantum noise effect should be enhanced by a classical randomness. 
So Alice and Bob in Y-00 share a secret key $K$. The key is stretched by a linear feedback shift register:LFSR as a pseudo random number generator to $K'$. The length of the initial key is $|K|=100 \sim 1000$, and the length of the running key is 
$|K'|\sim 2^{|K|}$.
The data bit is modulated by $M$-ary keying driven by random decimal number generated from the block :$K'/\log_2 M=\bar{K}'=(k_1,k_2,\dots)$ of pseudo random number with the seed secret key $K$. 
The $M$-ary keying has $M$ different basis based on 2$M$ coherent states. So the data bit is mapped into one of 2$M$ coherent states randomly.
A quantum state sequence emitted from the transmitter is as follows:
\begin{equation}
|\Psi \rangle =|\alpha_i \rangle_1 |\alpha_j \rangle_2 
|\alpha_k \rangle_3 \dots 
\end{equation}
where $|\alpha_i \rangle$ is one of 2$M$ coherent states, and $i,j,k\in {\cal{M}}=(1 \sim 2M)$. This sequence is one sequence  in a set of the $F$ sequences: 
\begin{equation}
F=(2M)^{2^{|K|}/\log_2{M}}
\end{equation}
That is, the density operator of Eve is
\begin{equation}
\rho_T =\sum^{F}_{l=1}\frac{1}{F}|\Psi_l \rangle \langle \Psi_l |
\end{equation}
The processing to break Y-00 is done by physical measurements of quantum state sequences. So we have to take not only Eve's power of computation but also ability of physical implementation into account. We describe several physical attacks on Y-00 in the following sections.

\subsubsection{Method of quantum optimum measurement}
{\bf (A)} Quantum individual processing\\
\\
All quantum state sequences have certain amounts of correlations  from PRN as running key,  because the running key is not a true random number. Let us assume that Eve employs individual quantum measurement for each state in the sequence, and classical processing for measurement results. If the state is orthogonal states, then Eve can get ciphertext without error. So the security comes from only the key uncertainty as classical randomness. However, in Y-00, $M$-ary scheme provides a set of non-orthogonal states.
Let us see what is the principle of security of Y-00. By introducing the classical randomness by secret key and pseudo random number generator, the performances of quantum measurements become as follows:
\begin{itemize}
\item[\rm(i)]  Ciphertext-only attack on data: quantum detection  is binary pure state signals for Bob, and  binary mixed states for Eve.
\item[\rm(ii)] Known/chosen plaintext attack, and Ciphertext-only attack on key: quantum detection  is  binary pure signals for Bob, and   M-ary pure states for Eve.
\end{itemize}
That is, the limitation for accuracy of measurement of Bob is given by Helstrom bound as follows[7]:
\begin{equation}
P_e= 1/2\min_{\Pi}(Tr\rho_1\Pi_0 + Tr\rho_{0}\Pi_1)\ll \frac{1}{2}
\end{equation}
On the other hand, Eve does not know the key and running key. So her  density operators for information bits become "mixed state".
 For example, in the case of ciphertext only quantum individual attack, they are
\begin{equation}
\rho_0 = \sum q_j|\alpha_j\rangle\langle \alpha_j|, \quad 
\rho_1 = \sum q_k|\alpha_k\rangle\langle \alpha_k|
\end{equation}
The probability $p_i$ depends on the statistics of the data, and $q_j$, $q_k$ depend on the pseudo random number with $j$, and  $k$ being even and odd number. Eve has to extract the  data from the quantum system with mixed states, and her error probability is also given by Helstrom bound for mixed states.
\begin{equation}
\bar{P}_e= \min_{\Pi}(p_1Tr\rho_1\Pi_0 + p_0Tr\rho_0\Pi_1)\sim \frac{1}{2}
\end{equation}
Thus, the error probability of Eve becomes $ 1/2$ from the appropriate choice of the number $M$, signal energy, and overlap selection keying(OSK)[17, 18]. It means that Eve's data $Y_E$ is completely inaccurate.
This is equivalent to one time pad[17].

On the ciphertext only attack on key and known/chosen plaintext attack, the best way for Eve is to detect $M$ basis based on 2$M$ coherent states. In this case, the limitation for accuracy of Eve's data is also given by the minimax quantum detection[19] of 2$M$ pure coherent states for ciphertext only attack on key, and $M$ for known/chosen plaintext attack on key. As a result, the measured data on the running key involve unavoidable error given by 
\begin{equation}
P_e = \max_{p_i}\min_{\Pi} (1 - \sum p_iTr \rho_i\Pi_i)
\end{equation}
For the basic model of Y-00, the above formula gives $P_e=0.975$ when $2M=2047,<n>=100$, and $P_e=0.755$ when $2M=2047,<n>=10000$[20,21].\\ 
\\
{\bf (B)} Quantum collective processing\\
\\
Here we are concerned with known/chosen plaintext attack. {\it Let us assume a very strong condition. That is, Eve can inserts all zero bit sequence as the full length plaintext in one period}.
In order to clarify the essential point, we first neglect the PRNG, and we employ only an initial secret key with length $|K|$ which is selected from a true random number.
The initial key is divided by $\log_2 M$, and the number of slot is $L=|K|/\log_2 M$. For known/chosen plaintext attack, the number of quantum states for Eve are $M$. So the total number of sequences is $M^{|K|/\log_2 M}=2^{|K|}$. 
 Eve can measure directly the state sequences by means of the collective measurement scheme in order to get the key.
The  density operators and her detection operators are described on a tensor product Hilbert space of 
$L=|K|/\log_2 M$. That is, 
\begin{equation}
{\bf{\rho}_n} = \rho_{i_{1}}\otimes \rho_{i_{2}} \otimes \rho_{i_{3}}\otimes \dots 
\end{equation}
\begin{equation}
{\bf P_n} = p_{i_{1}}p_{i_{2}}p_{i_{3}} \dots
\end{equation}
where $i=(1,2,\dots M)$, 
${\bf{\rho}_n}\in {\cal H}^{\otimes L}$, 
and $\bf{\Pi}_n \ge 0, \sum{\bf{\Pi}_n}$$=\otimes I_i$, and where $\bf n$ is $\{1,2, \dots 2^{|K|}\}$. Then we have the following problem.
\begin{equation}
\min \max {\bf P_e}= \min_{\bf{\Pi}_n}\max_{\bf P_n}(1- \sum {\bf P_n }Tr {\bf \rho_n \Pi_n})
\end{equation}
Fortunately, we can prove the following theorem.\\
\\
{\bf Theorem 3}.  {\it The optimum measurement of the collective quantum measurement for all sequences is individual quantum measurement for all slots:
\begin{equation}
{\bf{\Pi}_n}=\Pi_{i_{1}}\otimes \Pi_{i_{2}} \otimes \Pi_{i_{3}}\otimes \dots 
\end{equation}
and the success probability is }
\begin{equation}
{\bf P_D}= (\sum p_i Tr \rho_i \Pi_i)^L
\end{equation}
If the quantum states are orthogonal, then the success probability is 1, and Eve gets exact key by known plaintext attack in the first period.  However, we can design ${\bf P_D} << 1$ by using coherent states. So Y-00 has a potential to achieve $H(K|Y_E, X)>0$.

Here we shall discuss a situation of the out of framework of the cryptanalysis, but it is interesting problem. Since the key is fixed in Y-00, there is a period at $2^{|K|}/\log_2 M$ bits.
Let us assume that Eve can get the transmitter and try many times known/chosen full length plaintext attack.  As a result, she can try $J$ times known/chosen full length plaintext attack.
The success probability is
\begin{equation}
{\bf {\bar P}_D}= 1-(1-{\bf P_D})^J
\end{equation}
Thus, to determine the true key uniquely, she needs the infinite trial when ${\bf P_D} << 1$.
In the case of Y-00 using PRNG, the situation is complicated, but the essential point may be the same one.

\subsubsection{Method of collective measurement with all key}
Since the length of PRN is $2^{|K|}$, the transmitter can 
send $2^{|K|}/\log_2 M$ bits in the first communication.
Eve can insert her known/chosen plaintext into transmitter and measure the quantum state sequences by quantum receivers with all kinds of key. If Eve wants to get the true key, then Eve needs $2^{|K|}$ copies. But it is not allowed by no cloning theorem. Although  Eve can try a beam splitter attack so called {\bf Lo-Ko attack}, it has been shown that this attack does not work[6, 17], and it has no effect for the security analysis.

Let us discuss collective unambiguous quantum measurement attack. Eve can insert her known plaintext into the data port of the transmitter. She can prepare the unambiguous state discrimination:${\bf \Pi_{un}}$ which can apply to $2^{|K|}$ quantum state sequences.  
One of quantum state sequences of the set is transmitted from Alice. Eve will measure it by her unambiguous measurement. The success probability is evaluated by an exact calculation and also the following theorem.\\
\\
{\bf Theorem 4}. 
{\it The upper bound of average success probability in unambiguous measurement is given by the  quantum optimum solution in quantum detection theory for the same state ensemble.}\\
\\
The unambiguous state discrimination(USD) for $M$ symmetric coherent states is formulated by  A.Chefles and S.M.Barnett[22], and S.J.van Enk[23].
The success probability is given by the following formula.
\begin{equation}
P_D= N \min_{k=1,2,3, \dots, N} |c_k|^2
\end{equation}
where
\begin{equation}
|c_k|^2= \frac{1}{N}\sum^{N}_{j=1}e^{2\pi ijk/N}e^{|\alpha|^2(e^{2\pi ij/N}-1)}
\end{equation}
In fact, in the case of individual measurement, the unambiguous state discrimination(USD) on $M$=2000 symmetric coherent states with $(<n>=10000)$ is
\begin{equation}
P_D(USD) \sim 3{\rm x}10^{-12} <5{\rm x}10^{-4}=\frac{1}{M} < P_D(Bayes)\sim 2{\rm x}10^{-1}
\end{equation}
In addition, the success probability for collective USD is given by
\begin{equation}
{\bf P_D}(USD) < 2^{-|K|} < {\bf P_D}(Bayes)
\end{equation}
That is, the probability is less than that pure guessing.

Let us discuss again the  known/chosen full length plaintext attack of unlimited repetition, though this is impractical.
At the first trial, Eve measures the quantum state sequence by quantum collective measurement with the key $K_1$, at the second trial, Eve measures with $K_2$, and so on. She knows the plaintext. So when the output is the plaintext, the measurement has the true key.
Thus, in order that Eve gets the true key, she needs $2^{|K|}$ times known/chosen plaintext attack based on a full bit length 
$2^{|K|}/\log_2 M$. This may be only one method to break Y-00.
However, if the legitimate users change the key sometimes as usual(for example, once a year), then this attack has no meanings.
This type of discussion belongs to a key management. If Y-00 cannot be broken by any methods except for the above situation, then Y-00 is indeed the ultimate cipher.

\subsubsection{General properties}
The basis of the security is a combination of key uncertainty(classical randomness) and quantum noise. {\bf A role of classical randomness: secret key (or running key) is to make a difference of quantum measurement performance}. 
Thus, we can say that Y-00 is a randomized encryption cipher with no loss of bandwidth and with high speed randomization done by quantum noise of coherent state.
However, we should emphasize that an appropriate design is necessary to realize meaningful security. 

In the conventional theory, we have $H(X|Y_E, R_M) \le H(K)$  known as Shannon bound for ciphertext only attack on data, and $H(K|Y_E,R_M)\ge 0$ for ciphertext only attack on key which is relevant with "unicity distance", where $X$ is data sequence, $Y_E$ is Eve's data, and $R_M$ is public mathematical randomization, respectively. In addition, for known/chosen  plaintext attack, we have $H(K|X, Y_E, R_M) =0$ which means a computational complexity based security.

However, our goal is to show the following performance. 
In the cipher-text only attack on data, Y-00 may exceed the classical Shannon limit in the cryptography, even we use a system with $H(K) << H(X)$. That is,
\begin{equation}
H(X|Y_E, R_M, R_P) > H(K)
\end{equation}
where $X$ is information data, $Y_E$ is ciphertext which is  "measured value" of Eve, $R_P$ is physical randomization, and $K$ is initial secret key. 
For known/chosen plaintext attack, Y-00 has
\begin{equation}
H(K|Y_E,R_M, R_P, X) > 0
\end{equation}
which corresponds to information theoretic security. If one has $H(K|Y_E,R_M, R_P, X) =H(K)$, then it is perfect key security. 
These are not realized by the conventional symmetric key cipher.  It means that Y-00 has a potential to break the limitation of the conventional cryptography theory based on quantum communication theory. The proof for the general attacks will be shown in the subsequent papers based on the results in this paper.

\subsection{Quantum key generation}
Y-00 is also applicable to key generation. Let us introduce its basic concept.
In this case, data is a true random number sequence. So there is no criterion like known plaintext attack.  The use of shared secret key between Alice and Bob that determine the quantum states generated for the data bit sequences in a detection/coding scheme gives them a better error performance  over Eve who does not know $K$.
Based on the this principle, the general conditions for key generation were discussed by Yuen[6]. 
As a result, the condition for secure key generation is 
\begin{equation}
H(X_A|Y_E, K) > H(X_A|Y_B)
\end{equation}
where $Y_B$ is Bob's observation with knowledge of the seed key. 
A concrete implementation is coherent pulse position modulation scheme[6].

\section{Signal design for experiments}
In the phase modulation scheme(PSK), the coherent states are described by positions on a circle in the phase space representation. The radius corresponds to the amplitude or average photon number per pulse at the transmitter. The positions on the circle correspond to phase information of the light wave. If the number of basis is $M$, then the signal distance between  neighbor states is about $\Delta_{PM} =\frac{2\pi |\alpha|}{2M}$.
The uncertainty of coherent state is described by two dimensional Gaussian distribution with mean$=|\alpha|$ and variance$=1/4$.
In the practical sense, we can design the number of basis which satisfies 
\begin{equation}
P_e(i, i+1) = \frac{1}{2} - \frac{1}{\sqrt{2\pi}}\int_{0}^{t_0}\exp(-t^2/2) dt
=0.2 \sim 0.5
\end{equation}
where $t_0= \Delta_{PM}/2=\frac{\pi |\alpha|}{2M}$. This corresponds to the error probability between neighbor states.
But it is not real error probability for Eve. 
The real error probability of Eve depends on her strategy and quantum measurement scheme. On the other hand, for amplitude or intensity modulation scheme(ASK), the conditions for parameters are $\Delta_{AM} =\frac{|\alpha_{max}-\alpha_{min}|}{2M}$, $t_0= \Delta_{AM}/2$, 
and $|\alpha_{min}|^2 > \frac{1}{\kappa}$, where $\kappa$ is efficiency of permeability of attenuation channel.

\section{Conclusion}
We have given a rigorous interpretation of an origin of the security of Y-00. Although the classical randomness as key uncertainty is essential in Y-00, the security is given by quantum randomness. 
Directions to security analysis have been given in this paper and [6, 17]. The proof for the general attacks will be shown in the subsequent papers. However, in practical use, exponential complexity for known plaintext attack may be enough, which was already demonstrated by experiments.

\section*{Acknowledgment}
OH is grateful to H.P.Yuen, P.Kumar and E.Corndorf of Northwestern University for discussions.



\end{document}